\documentstyle[12pt,axodraw,epsfig]{article}          
\makeatletter  

\@addtoreset{equation}{section} \makeatother

\newcommand{\hs}{\hspace*{0.5cm}}

\newcommand{\be}{\begin{equation}}
\newcommand{\ee}{\end{equation}}
\newcommand{\bea}{\begin{eqnarray}}
\newcommand{\eea}{\end{eqnarray}}
\newcommand{\ben}{\begin{enumerate}}
\newcommand{\een}{\end{enumerate}}
\newcommand{\nn}{\nonumber}
\newcommand{\crn}{\nonumber \\}

\newcommand{\la}{\lambda}

\newcommand{\ga}{\gamma}

\newcommand{\fr}{\frac}
\newcommand{\bc}{\begin{center}}
\newcommand{\ec}{\end{center}}

\def\lappeq{\mathrel{\rlap{\raise.5ex\hbox{$<$}}
{\lower.5ex\hbox{$\sim$}}}}
\begin{document}

\bc {\Large Neutrino Masses in  Supersymmetric Economical \\
 $\mbox{SU}(3)_{C}\otimes \mbox{SU}(3)_{L} \otimes \mbox{U}(1)_{X}$ Model}\\
\vspace*{1cm}

{\bf P. V. Dong$^a$, D. T. Huong$^{a}$, M. C. Rodriguez$^b$}
and {\bf H. N. Long$^a$}\\

\vspace*{0.5cm}

$^a$ {\it Institute of Physics, VAST, P. O. Box 429, Bo Ho, Hanoi
10000, Vietnam}\\

 $^b$ {\it Universidade Federal do Rio Grande,
 Instituto de Matem\'atica, Estat\'\i stica e F\'\i sica,\\
 Av. It\'alia, km 8, Campus Carreiros,
 96201-900 Rio Grande, RS,  Brazil}

\ec

\begin{abstract}
The $R$-symmetry formalism is applied for the supersymmetric
economical $\mbox{SU}(3)_{C}\otimes \mbox{SU}(3)_{L} \otimes
\mbox{U}(1)_{X}$ (3-3-1) model.
The generalization of the minimal supersymmetric standard model
relation among $R$-parity, spin and matter parity is derived, and
discrete symmetries for the proton stability in this model are
imposed. We show that in such a case it is able to give leptons
masses at
just the tree level.
 A simple mechanism for the mass generation of the
neutrinos is explored. With the new $R$-parity, the neutral
fermions get mass matrix with two distinct sectors: one light
which is identified with neutrino mass matrix, another heavy one
which is identified with neutralinos one. The similar situation
exists in the charged fermion sector. Some phenomenological
consequences such as proton stability, neutrinoless double beta
decays are discussed.
\end{abstract}


\section{Introduction}

Although the Standard Model (SM) gives very good results in
explaining the observed properties of the charged fermions, it is
unlikely to be the ultimate theory. It maintains the masslessness
of the neutrinos to all orders in perturbation theory, and even
after non-pertubative effects are included. The recent
groundbreaking discovery of nonzero neutrino masses and
oscillations \cite{superk} has put massive neutrinos as one of
evidences on physics beyond the SM.

The Super-Kamiokande experiments on the atmospheric neutrino
oscillations have indicated to the difference of the squared
masses and the mixing angle with fair accuracy~\cite{kam,sno}
\begin{eqnarray}
\label{eqn::atm}
\Delta m^2_{\mathrm{atm}}  & = & 1.3 \div 3.0 \times 10^{-3}  {\rm eV ^2}, \\
\sin^2 2 \theta_{\mathrm{atm}}  & > & 0.9.
\end{eqnarray}
While, those from the combined fit of the solar and reactor
neutrino data point to
\begin{eqnarray}
\label{eqn::solar}
\Delta m^2_{\odot} ~ & = &
8.0^{+ 0.6}_{- 0.4} \times 10^{-5} ~ \rm{ eV^2}, \\
\tan ^2 \theta_{\odot} ~ & = & 0.45^{+0.09}_{-0.07}.
\end{eqnarray}
Since the data provide only the information about the differences
in $m_{\nu}^2$, the neutrino mass pattern can be either almost
degenerate or hierarchical. Among the hierarchical possibilities,
there are two types of normal and inverted hierarchies. In the
literature, most of the cases explore normal hierarchical one in
each. In this paper, we will mention on a supersymmetric model
which naturally gives rise to three pseudo-Dirac neutrinos with an
inverted hierarchical mass pattern.

The gauge symmetry of the SM as well as those of many extensional
models by themselves fix only the gauge bosons. The fermions and
Higgs contents have to be chosen somewhat arbitrarily. In the SM,
these choices are made in such a way that the neutrinos are
massless as mentioned. However, there are other choices based on
the SM symmetry that neutrinos become massive. We know these from
the popular seesaw \cite{seesaw} and radiative \cite{rad} models.
Particularly, the models based on the $\mbox{SU}(3)_C\otimes
\mbox{SU}(3)_L \otimes \mbox{U}(1)_X$ gauge unification group
\cite{ppf,flt,recent}, called 3-3-1 models, give more stricter
fermion contents. Indeed, only three fermion generations are
acquired as a result of the anomaly cancelation and the condition
of QCD asymptotic freedom. The arbitrariness in this case are only
behind which SM singlets put in the bottoms of the lepton
triplets? In some scenarios, exotic leptons may exist in the
singlets. Result of this is quite similar the case of the SM
neutrinos. As a fact, the mechanisms of the Zee's type \cite{rad}
for neutrino masses arise which been explored in Ref.\cite{yakip}.

Forbidding the exotic leptons, there are two main versions of the
3-3-1 models as far as minimal lepton sectors is concerned. In one
of them \cite{ppf} the three known left-handed lepton components
for each generation are associated to three $\mbox{SU}(3)_L$
triplets as $(\nu_l,l,l^c)_L$, in which $l^c_L$ is related to the
right-handed isospin singlet of the charged lepton $l$ in the SM.
No extra leptons are needed and therefore it calls that a minimal
3-3-1 model. In the variant model \cite{flt} three
$\mbox{SU}(3)_L$ lepton triplets are of the form $(\nu_l, l,
\nu_l^c)_L$, where $\nu_l^c$ is related to the right-handed
component of the neutrino field $\nu_l$, thus called a model with
the right-handed neutrinos. This kind of the 3-3-1 models requires
only a more economical Higgs sector for breaking the gauge
symmetry and generating the fermion masses. It is interesting to
note that two Higgs triplets of this model have the same
$\mathrm{U}(1)_X$ charges with two neutral components at their top
and bottom.  Allowing these neutral components vacuum expectation
values (VEVs) we can reduce number of Higgs triplets to be two.
Therefore we have a resulting 3-3-1 model with two Higgs triplets
\cite{haihiggs}. As a result, the dynamical symmetry breaking also
affects lepton number. Hence it follows that the lepton number is
also broken spontaneously at a high scale of energy. Note that the
mentioned model contains very important advantage, namely, there
is no new parameter, but it contains very simple Higgs sector,
therefore the significant number of free parameters is reduced. To
mark the minimal content of the Higgs sector, this version that
includes right-handed neutrinos is called the {\it economical
3-3-1 model}.

Among the new gauge bosons in this model, the neutral
non-Hermitian bilepton field $X^0$ may give promising signature in
accelerator experiments and may be also the source of neutrino
oscillations~\cite{til}. In the current paper, the neutrinos of
the 3-3-1 model with right-handed neutrinos is a subject for
extended study.

The 3-3-1 model with right-handed neutrinos gives the tree level
neutrino mass spectrum with three Dirac fermions, one massless and
two degenerate in mass \cite{changlong}. This is clearly not
realistic under the experimental data. However, this pattern may
be severely changed by quantum effects and gives rise to an
inverted hierarchy mass pattern. This is a specific feature of the
3-3-1 model with right-handed neutrinos which was considered in
Ref.\cite{changlong} (see also \cite{dias}), but such effects
exist in the very high level of the loop corrections.

The outline of this work is as follows. In Sec.
\ref{sec:rparitysusy3312} we define the $R$-charge in our model in
order to get similar results as in the Minimal Supersymmetric
Standard Model (MSSM). While in Sec. \ref{sec:neutrinosmassterms}
we impose another discrete symmetry that allow neutrino masses but
forbid the proton decay and the neutron-antineutron oscillation.
In Sec. \ref{sec:fermionmasses} we calculate the fermion masses in
our model, then we present some phenomenological discussion of
this model.Our conclusions are found in the last section. In
Appendix \ref{neumatel}, we present the mass matrix elements of
the neutral fermions.

\section{Discrete R-parity in the supersymmetric
economical 3-3-1 model (SUSYECO331).}
\label{sec:rparitysusy3312}

In the supersymmetric 3-3-1 model with right-handed neutrinos
(SUSY331RN)~\cite{susy3312}, the $R$-parity was already studied
and we have shown that if $R$-symmetry is broken, the simple
mechanism for the neutrinos  mass can be constructed
~\cite{Dong:2006vk}. This mechanism produces the neutrinos mass
which is in  agreement with the experimental data.

The supersymmetric extension of the economical 3-3-1 model
(SUSYECO331) was presented in~\cite{Dong:2007qc}. The fermionic
content of SUSYECO331 is the following: the left-handed fermions
are in the triplets/antitriplets under the $SU(3)_L$ group,
namely,  the usual leptons are the  triplets
$L_{i}=(\nu_{i},l_{i}, \nu^{c}_{i})_L \sim({\bf1},{\bf3},-1/3)$,
$i=1,2,3$; while in the quark sector,  we have two families in the
antitriplets $Q_{\alpha L}=(d_{\alpha},u_{\alpha},D_{\alpha})
\sim({\bf3},{\bf3}^*,0)$, $\alpha=1,\ 2$, and one family in the
triplet $Q_{3L}=(u_{3},d_{3},T)\sim({\bf3},{\bf3},1/3)$. The
right-handed components are in the singlets  under the $SU(3)_L$
group: $l^{c}_{i}\sim({\bf1},{\bf1},1)$, $u^{c}_{i} \sim({\bf3}^*,
{\bf1},-2/3)$, $d^{c}_{i}\sim({\bf3}^*,{\bf1},1/3)$, which are
similar to those in the SM. In addition,  the exotic quarks
transform as $T^{c}\sim({\bf3}^*, {\bf1},-2/3),\
D^{c}_{\alpha}\sim({\bf3}^*,{\bf1},1/3)$.

The scalar content is minimally formed by two Higgs triplets:
$\chi=(\chi^{0}_{1},\chi^{-},\chi^{0}_{2})^T\sim({\bf1},{\bf3},-1/3)$
and
$\rho=(\rho^{+}_{1},\rho^{0},\rho^{+}_{2})^T\sim({\bf1},{\bf3},2/3)$.
In order to cancel chiral anomalies in the SUSYECO331 model, we
have  to introduce the followings scalar Higgs triplets
$\chi^{\prime}=(\chi^{\prime 0}_{1}, \chi^{\prime +},\chi^{\prime
0}_{2})^T\sim({\bf1},{\bf3}^*,1/3)$ and
$\rho^{\prime}=(\rho^{\prime -}_{1},\rho^{\prime 0},\rho^{\prime
-}_{2})^T\sim({\bf1},{\bf3}^*,-2/3)$. In this model, the VEVs are
defined by
\begin{eqnarray}
 \sqrt{2} \langle\chi\rangle^T &=& \left(u, 0, w\right), \hs \sqrt{2}
 \langle\chi^\prime\rangle^T = \left(u^\prime,  0,
 w^\prime\right),\\
\sqrt{2}  \langle\rho\rangle^T &=& \left( 0, v, 0 \right), \hs
\sqrt{2} \langle\rho^\prime\rangle^T = \left( 0, v^\prime,  0
\right).
\label{vev}
\end{eqnarray}
The VEVs $w$ and $w^\prime$ are responsible for the first step of
the symmetry breaking, while $u,\ u^\prime$ and $v,\ v^\prime$ are
responsible for the second one. Therefore, they have to satisfy
the constraints:
\begin{equation}
 u,\ u^\prime,\ v,\ v^\prime
\ll w,\ w^\prime. \label{contraint}
\end{equation}
The complete set of fields and the full lagrangian of SUSYECO331
are given in Ref.~\cite{Dong:2007qc}. The most general
superpotential is given by:
\begin{equation}
W= \frac{W_{2}}{2}+ \frac{W_{3}}{3},
\end{equation}
where
\begin{eqnarray}
W_{2}&=& \mu_{0i}\hat{L}_{iL} \hat{ \chi}^{\prime}+ \mu_{ \chi}
\hat{ \chi} \hat{ \chi}^{\prime}+
 \mu_{ \rho} \hat{ \rho} \hat{ \rho}^{\prime},
 \label{w2}
\end{eqnarray}
and
\begin{eqnarray}
W_{3}&=& \ga_{ab} \hat{L}_{aL} \hat{ \rho}^{\prime}
\hat{l}^{c}_{bL}+ \la_{a} \epsilon \hat{L}_{aL} \hat{\chi}
\hat{\rho}+ \la^\prime_{ab} \epsilon \hat{L}_{aL} \hat{L}_{bL}
\hat{\rho} \nonumber \\
&+& \kappa_{i} \hat{Q}_{1L} \hat{\chi}^{\prime} \hat{u}^{c}_{iL}+
\kappa^\prime \hat{Q}_{1L} \hat{\chi}^{\prime} \hat{u}^{\prime
c}_L+ \vartheta_{i}\hat{Q}_{1L} \hat{\rho}^{\prime}
\hat{d}^{c}_{iL} \nonumber \\
&+& \vartheta^\prime_{ \alpha}\hat{Q}_{1L} \hat{\rho}^{\prime}
\hat{d}^{\prime c}_{\alpha L} + \pi_{ \alpha i} \hat{Q}_{\alpha
L}\hat{\rho}\hat{u}^{c}_{iL} +\pi_{\alpha}^{\prime}
\hat{Q}_{\alpha L}\hat{\rho}\hat{u}^{\prime c}_{L} \nonumber \\
&+& \Pi_{\alpha i} \hat{Q}_{\alpha L} \hat{\chi} \hat{d}^{c}_{iL}
+ \Pi^\prime_{\alpha \beta} \hat{Q}_{\alpha L} \hat{\chi}
\hat{d}^{\prime c}_{\beta L}+ \epsilon
f_{\alpha\beta\gamma}\hat{Q}_{\alpha L} \hat{Q}_{\beta L}
\hat{Q}_{\gamma L} \crn &+& \xi_{1i \beta j} \hat{d}^{c}_{iL}
\hat{d}^{\prime c}_{\beta L} \hat{u}^{c}_{j L}+ \xi_{2i \beta }
\hat{d}^{c}_{i L} \hat{d}^{\prime c}_{\beta L} \hat{u}^{\prime
c}_{L}+ \xi_{3ijk}
\hat{d}^{c}_{iL} \hat{d}^{c}_{jL} \hat{u}^{c}_{k L} \nonumber \\
&+& \xi_{4ij} \hat{d}^{c}_{i L} \hat{d}^{c}_{jL} \hat{u}^{\prime
c}_{L}+ \xi_{5 \alpha \beta i} \hat{d}^{\prime c}_{\alpha L}
\hat{d}^{\prime c}_{\beta L} \hat{u}^{c}_{iL} + \xi_{6 \alpha
\beta} \hat{d}^{\prime c}_{\alpha L}\hat{d}^{\prime c}_{\beta L}
\hat{u}^{\prime c}_{L} \crn &+& \xi_{a \alpha j}\hat{L}_{aL}
\hat{Q}_{\alpha L} \hat{d}^{c}_{jL}+ \xi^\prime_{a\alpha
\beta}\hat{L}_{aL} \hat{Q}_{\alpha L} \hat{d}^{\prime c}_{\beta
L}. \label{w3}
\end{eqnarray}

The coefficients $\mu_{0a}, \mu_{\rho}$ and $\mu_{\chi}$ have mass
dimension and can be complex variables \cite{mssm}, while all
coefficients in $W_{3}$ are dimensionless, and $\la^\prime_{ab}= -
\la^\prime_{ba}$.

Let us impose the $R$-parity as the same as that of the minimal
supersymmetric standard model. In this case, we have to choose the
following $R$-charges \bea n_{\chi}&=&n_{\chi^{\prime}}=n_{\rho}=
n_{\rho^{\prime}}=0, \crn n_{L}&=&n_{Q_{\alpha}}=n_{Q_{3}}=1/2,
\crn n_{l}&=&n_{u}=n_{d}=n_{T}=n_{D}=-1/2. \label{rdiscsusy331rn}
\eea The superpotential satisfying the above $R$-parity
conservation is written as
\bea W_{RC}&=& \frac{\mu_{ \chi}}{2}
\hat{ \chi} \hat{\chi}^{\prime}+ \frac{\mu_{ \rho}}{2} \hat{ \rho}
\hat{ \rho}^{\prime}+ \frac{1}{3} \left[ \lambda_{1ij} \hat{L}_{i}
\hat{ \rho}^{\prime} \hat{l}^{c}_{j}+ \kappa_{1i} \hat{Q}_{3}
\hat{\eta}^{\prime} \hat{u}^{c}_{i}+ \kappa_{1}^{\prime}
\hat{Q}_{3} \hat{\eta}^{\prime} \hat{T}^{c}+ \kappa_{2i} \hat{Q}_{3}
\hat{\chi}^{\prime} \hat{u}^{c}_{i}+ \kappa_{2}^{\prime}
\hat{Q}_{3} \hat{\chi}^{\prime} \hat{T}^{c} \right. \crn &+& \left.
\kappa_{3\alpha i} \hat{Q}_{\alpha} \hat{\eta} \hat{d}^{c}_{i} +
\kappa_{3\alpha \beta}^{\prime} \hat{Q}_{\alpha} \hat{\eta}
\hat{D}^{c}_{\beta}  + \kappa_{4\alpha i}
\hat{Q}_{\alpha}\hat{\rho}\hat{u}^{c}_{i}+
\kappa_{4\alpha}^{\prime} \hat{Q}_{\alpha}\hat{\rho} \hat{T}^{c}+
\kappa_{5i}\hat{Q}_{3} \hat{\rho}^{\prime} \hat{d}^{c}_{i}+
\kappa_{5 \beta}^{\prime}\hat{Q}_{3} \hat{\rho}^{\prime}
\hat{D}^{c}_{\beta}\right. \crn & +& \left.  \kappa_{6\alpha i}
\hat{Q}_{\alpha} \hat{\chi} \hat{d}^{c}_{i}+ \kappa_{6\alpha
\beta}^{\prime} \hat{Q}_{\alpha} \hat{\chi} \hat{D}^{c}_{\beta}
\right]. \label{rpartseco331} \eea
 With this superpotential, we have
shown that ~\cite{Dong:2007qc} the boson,  Higgs sectors and the
fermion one gain masses.

Thus, the $R$-parity in this model, as in the SUSY331RN, can be
re-expressed via the spin $S$, new charges $\mathcal{L}$ and
$\mathcal{B}$ in terms of
\cite{Dong:2006vk}
\begin{equation}
\hbox{\emph{R}-parity}=(-1)^{2S}(-1)^{3({\cal B}+{\cal L})},
\end{equation}
where the charges $ {\cal B}$ and ${\cal L}$ for the multiplets
are defined as follows \cite{changlong}
\begin{equation}
\begin{array}{|c|c|c|c|c|c|}
\hline
  Triplet & L & Q_{3} & \chi &  \rho \\
  \hline
  {\cal B} \,\  charge & 0 & \frac{1}{3} & 0 & 0 \\ \hline
  {\cal L} \,\  charge & \frac{1}{3} &
  - \frac{2}{3} & \frac{4}{3} & - \frac{2}{3} \\ \hline
\end{array}
\end{equation}
\begin{equation}
\begin{array}{|c|c|c|c|c|}
\hline
  Anti-triplet & Q_{\alpha} & \chi^{\prime} &  \rho^{\prime} \\
  \hline
  {\cal B} \,\  charge & \frac{1}{3} & 0 & 0 \\ \hline
  {\cal L} \,\  charge & \frac{2}{3} & - \frac{4}{3} & \frac{2}{3} \\ \hline
\end{array}
\end{equation}
\begin{equation}
\begin{array}{|c|c|c|c|c|c|}
\hline
  Singlet & l^{c} & u^{c} & d^{c} & T^{c} & D^{c} \\
  \hline
  {\cal B} \,\ charge & 0 & -\frac{1}{3} & -\frac{1}{3} &
  -\frac{1}{3} & -\frac{1}{3} \\ \hline
  {\cal L} \,\ charge & -1 & 0 &  0 & 2 & -2 \\ \hline
\end{array}
\end{equation}

From the superpotential given in Eq.(\ref{rpartseco331}), it is
easy to see that the charged leptons gain mass through the term
\be -\frac{\lambda_{1ij}}{3}L_{i} \rho^{\prime}l^{c}_{j}+hc.
\label{shklcl} \ee
Their mass matrix, see Eq.(\ref{vev}), is given by
\begin{equation}
\frac{v^{\prime}}{3 \sqrt{2}} \lambda_{1ij}.
\end{equation}
Note that there is only VEV of $\rho^{\prime}$ given the
charged leptons masses.

Unfortunately, as in the MSSM  case,  due to conservation of the
$R$-parity defined in Eq.(\ref{rdiscsusy331rn}), there are no term
which gives neutrinos masses. However,
looking at the superpotential of this model given in
Eq.(\ref{w3}), we see that there is a term
$\hat{L}_{i}\hat{L}_{j}\hat{\rho}$, which generates the following
\begin{equation}
\lambda_{3ij} \left( L_{i}L_{j}\rho +
L_{i}\tilde{L}_{j}\tilde{\rho}+ \tilde{L}_{i}L_{j}\tilde{\rho}
\right) \label{thh1}
\end{equation}
The first term  in (\ref{thh1}) generates  the following neutrino
mass matrix
\begin{equation}
\frac{v}{3 \sqrt{2}} \lambda_{3ij}.
\end{equation}
As shown  in  Ref.~\cite{Dong:2007qc}, the mass pattern of this
sector is $0$, $0$, $m_{\nu}$, $m_{\nu}$, $m_{\nu}$, $m_{\nu}$.
Note  that  in this case, we have two massless neutrinos.
Unfortunately, as in the non-symmetric version, the quantum
corrections at one loop level cannot  generate the realistic  mass
spectrum to the neutrinos. To get the realistic neutrino masses,
one have to introduce new physics scale or inflaton with mass
around the GUT scale \cite{nmass}.

In this article, we will explore a new mass mechanism to generate
neutrino masses at tree level for all neutrinos
and study the flavor violating processes, such as neutrinoless
double beta decay, which do not exist in our previous work.

\section{The discrete symmetry for
proton stability and neutrino masses in SUSYECO331}
 \label{sec:neutrinosmassterms}
In this section to get  neutrino mass and  impose  flavor
violating processes,  we  chose the new $R$ charge as follows
\bea
n_{L}&=&n_{\chi}= \frac{1}{2}, \crn
n_{\chi^{\prime}}&=&n_{u}=n_{T}=- \frac{1}{2}, \crn
n_{Q_{3}}&=&n_{\rho^{\prime}}=1, \ n_{d} = n_{D}=-2, \crn
n_{\rho}&=&-1, \ n_{Q_{\alpha}} = \frac{3}{2}, \ n_{l}=-
\frac{3}{2}. \label{Rnew}\eea This $R$-charge is different from
those presented in Ref.~\cite{Dong:2007qc}. We will show that in
this case, there exist some new phenomena, which are
 previously not allowed.

The terms under this symmetry are obtained by
\bea
W&=&W_{RC}+ \frac{1}{2} \mu_{0i}\hat{L}_{i}\hat{\chi}^{\prime} +
\frac{1}{3} \left( \lambda_{2i} \epsilon \hat{L}_{i} \hat{\chi}
\hat{\rho}+ \lambda_{3ij} \epsilon \hat{L}_{i} \hat{L}_{j}
\hat{\rho}+ \lambda^{\prime}_{\alpha
ij}\hat{Q}_{\alpha}\hat{L}_{i} \hat{d}^{c}_{j}+ \xi_{2 \alpha i
\beta}\hat{Q}_{\alpha}
\hat{L}_{i} \hat{D}^{c}_{\beta} \right), \nonumber \\
\label{neutrinomassesseco331} \eea where $W_{RC}$ is defined
in Eq.(\ref{rpartseco331}).  The superpotential given in
(\ref{neutrinomassesseco331}) will not only allow some interesting
flavor  violating processes
 but also  will simultaneously  give the nucleons a stability. As we will
show in next section, this superpotential will also generate
masses to all neutrinos in the model.

\section{Fermion masses}
\label{sec:fermionmasses}

The superpotential (\ref{neutrinomassesseco331}) provides us the
mixing between the leptons and higgsinos
as \be - \frac{\mu_{0i}}{2}L_{i}\tilde{\chi}^{\prime}-
\frac{\lambda_{2i}}{3}(L_{i}\tilde{\chi}\rho +
\tilde{\rho}L_{i}\chi)+ H.c. \ee

With the above terms, we get the mass matrices for the neutral and
charged fermions.
Diagonalizing these matrices we obtain the physical masses for the
fermions. Firstly, let us study the neutral fermions masses.

\subsection{Masses of the neutral fermions}

 In the basis $\Psi^0$ of the form
\bea \left( \nu_{1}\mbox{ }\nu_{2}\mbox{ } \nu_{3}\mbox{}
\nu^{c}_{1}\mbox{ } \nu^{c}_{2}\mbox{ } \nu^{c}_{3}\mbox{}
\tilde{\chi}^{0}_{1}\mbox{ } \tilde{\chi}^{\prime 0}_{1}\mbox{ }
\tilde{\chi}^{0}_{2}\mbox{ } \tilde{\chi}^{\prime 0}_{2}\mbox{ }
\tilde{\rho}^{0}\mbox{ } \tilde{\rho}^{\prime 0}\mbox{ }
\tilde{\mathcal{B}}\mbox{ } \tilde{\mathcal{W}}_{3}\mbox{ }
\tilde{\mathcal{W}}_{8}\mbox{ } \tilde{\mathcal{X}}\mbox{ }
\tilde{\mathcal{X}^*}\right),\nn \eea  the mass Lagrangian can be
written as follows \be - \frac{1}{2}(
\Psi^{0})^{T}Y^{0}\Psi^{0}+H.c.\label{mc1}\ee Here $Y^0$ is
symmetric matrix with the nonzero elements given in
Appendix.\ref{neumatel}, where the mass eigenstates are given by
\bea \tilde{ \chi}^{0}_{i}&=&N_{ij} \Psi^{0}_{j}, \,\ j=1, \cdots
,17. \label{emasneu}\eea The mass matrix of the neutral fermions
consists of three parts: (a) The first part $M_{\nu}$ is the
$6\times 6$ mass matrix of the neutrinos which belongs to the
SUSYECO331; (b) The second part $M_{N}$ is the $11\times 11$ mass
matrix of the neutralinos, which exists only in the
\emph{presented } supersymmetric version, has been  analyzed
in~\cite{Huong:2008ww}; (c) The last part $M_{\nu N}$ arises due
to mixing among the neutrinos and the neutral higgsinos. Thus, the
mass matrix
for the neutral fermions is signified as follows\be Y^0=\left(%
\begin{array}{cc}
  \left( M_{\nu} \right)_{6 \times 6} & \left( M_{\nu N} \right)_{11 \times 6} \\
  \left( M^T_{\nu N} \right)_{6 \times 11} & \left( M_{N} \right)_{11 \times 11} \\
\end{array}
\right), \label{neumatr} \ee where   matrices $M_{\nu}$ and
$M_{\nu N}$ are presented in Eq.(\ref{neumatr1}) and Eq.
(\ref{neumatr2}).

Let us  keep the mass constraints from astrophysics and cosmology
\cite{pdg} as well as being consistent with all the earlier
analysis \cite{lepmass}, the parameters in the mass matrix $M_N$
can be chosen as a typical example:
\bea
\mu_{\rho}&=&600\ \mathrm{GeV},\ \mu_{\chi}=700\ \mathrm{GeV},\crn
{\cal M}_{3}&=&{\cal M}_{8}=300 \mathrm{GeV},\
{\cal M}_{45}=400\ \mathrm{GeV},\nonumber \\
\mu_{01}&=&\mu_{02}=\mu_{03}=1\ \mathrm{GeV}, \nonumber \\
\lambda_{21}&=&\lambda_{22}=\lambda_{03}=1,\nonumber \\
\lambda_{312}&=&4 \cdot 10^{-11}, \ \lambda_{313}=5 \cdot 10^{-11}, \nonumber \\
\lambda_{321}&=&6 \cdot 10^{-11}, \ \lambda_{323}=7 \cdot 10^{-11}, \nonumber \\
\lambda_{331}&=&8 \cdot 10^{-11}, \ \lambda_{332}=9 \cdot 10^{-11}
,\label{para} \eea Here in this model, the Higgs
bosons' VEVs are fixed as follows\bea v_{\chi_1}&=&15
\mathrm{GeV},\ v_{\chi'_1}=10\ \mathrm{GeV}, \nonumber
\\
v_{\rho}&=&244.9\ \mathrm{GeV},\ v_{\rho'}=13\ \mathrm{GeV}, \crn
v_{\chi_2}&=&v_{\chi'_2}=1000\ \mathrm{GeV}, \label{vevs} \eea
and the value of $g$ is  given in Ref. \cite{pdg}.

Using the values given in Eqs.(\ref{para},\ref{vevs}), the
eigenvalues of fermion mass matrix  are obtained as
\begin{eqnarray}
m_{\tilde{\chi}^{0}_{17}}&=&-1282,27\ \mathrm{GeV}, \
m_{\tilde{\chi}^{0}_{16}}=1236,53\ \mathrm{GeV}, \
m_{\tilde{\chi}^{0}_{15}}=1041\ \mathrm{GeV}, \
m_{\tilde{\chi}^{0}_{14}}=-705,43\ \mathrm{GeV}, \nonumber \\
m_{\tilde{\chi}^{0}_{13}}&=&631,46\ \mathrm{GeV},
m_{\tilde{\chi}^{0}_{12}}=620,21\ \mathrm{GeV}, \
m_{\tilde{\chi}^{0}_{11}}=487,49\ \mathrm{GeV}, \
m_{\tilde{\chi}^{0}_{10}}=385,08\ \mathrm{GeV}, \nonumber \\
m_{\tilde{\chi}^{0}_{9}}&=&304,82\ \mathrm{GeV}, \
m_{\tilde{\chi}^{0}_{8}}=186,45\ \mathrm{GeV}, \
m_{\tilde{\chi}^{0}_{7}}=57,13\ \mathrm{GeV}, \
m_{\tilde{\chi}^{0}_{6}}=-0,103\ \mathrm{GeV}, \nonumber \\
m_{\tilde{\chi}^{0}_{5}}&=&-0,043\ \mathrm{GeV}, \
m_{\tilde{\chi}^{0}_{4}}=2,0415 \cdot 10^{-11}\ \mathrm{GeV}, \
m_{\tilde{\chi}^{0}_{3}}=2,0413 \cdot 10^{-11}\ \mathrm{GeV}, \nonumber \\
m_{\tilde{\chi}^{0}_{2}}&=&-2,0412 \cdot 10^{-11}\ \mathrm{GeV}, \
m_{\tilde{\chi}^{0}_{1}}=-2,0410 \cdot 10^{-11}\ \mathrm{GeV}.
\label{completeneutrinosmassespectrum}
\end{eqnarray}

In the Eq.(\ref{completeneutrinosmassespectrum}), there are some
\textit{negative} eigenvalues. In order to obtain the positive
mass, the eigenstates need to be redefined by the chiral
rotations.

Eq.(\ref{completeneutrinosmassespectrum}) shows that we have two
very distinct sector, one contains the light neutral fermions that
we will associate with the usual neutrinos in the SM  and the
other one contains the heavy neutralinos. The lightest neutralino
 mass equals to $57$ GeV and it is consistent with limits on
inelastic dark matter from ZEPLIN-III \cite{Akimov:2010vk}.

Using the values given in Eqs.(\ref{para},\ref{vevs}), the
eigenvalues of the mass matrix $\left( M_{\nu} \right)_{6 \times
6}$  are obtained as
\begin{eqnarray}
m_{\tilde{\chi}^{0}_{1}}&=&m_{\tilde{\chi}^{0}_{2}}=0\
\mathrm{GeV}, \crn
m_{\tilde{\chi}^{0}_{3}}&=&m_{\tilde{\chi}^{0}_{4}}=m_{\tilde{\chi}^{0}_{5}}=
m_{\tilde{\chi}^{0}_{6}}=2,125 \cdot 10^{-20}\ \mathrm{GeV}.
\label{neutrinosmassespectrum}
\end{eqnarray}

These values are smaller than that of the new mechanism given in
 (\ref{completeneutrinosmassespectrum}). On the other hand,
the eigenvalues of the matrix $\left( M_{N} \right)_{11 \times
11}$ are obtained by putting the numerical given in
Eqs.(\ref{para},\ref{vevs}) as follows
\begin{eqnarray}
m_{\tilde{\chi}^{0}_{7}}&=&1207\ \mathrm{GeV}, \
m_{\tilde{\chi}^{0}_{8}}=1143\ \mathrm{GeV}, \
m_{\tilde{\chi}^{0}_{9}}=1040\ \mathrm{GeV}, \
m_{\tilde{\chi}^{0}_{10}}=704,61\ \mathrm{GeV}, \crn
m_{\tilde{\chi}^{0}_{11}}&=&585,49\ \mathrm{GeV},
m_{\tilde{\chi}^{0}_{12}}=499,85\ \mathrm{GeV}, \
m_{\tilde{\chi}^{0}_{13}}=429,68\ \mathrm{GeV}, \
m_{\tilde{\chi}^{0}_{14}}=374,72\ \mathrm{GeV}, \crn
m_{\tilde{\chi}^{0}_{15}}&=&304,81\ \mathrm{GeV}, \
m_{\tilde{\chi}^{0}_{16}}=175,89\ \mathrm{GeV}, \
m_{\tilde{\chi}^{0}_{17}}=56,88\ \mathrm{GeV}.
\label{neutralinosmassespectrum}
\end{eqnarray}

These results can be understood as follows: Because of the
interference matrix $(M_{\nu N })_ {6 \times 11}$ between the
neutrino mass matrix and the neutralino  mass matrix, all
neutrinos gain mass at the tree level.  This change of the
neutrino mass spectrum is suitable to experiment data.
Thus  the neutrino mass spectrum in the   model under
consideration depends on the choice of   $R$-parity. Now we deal
with the charged fermions.

\subsection{Masses of the charged fermions}

 To write  mass matrix of the charged fermions, we will choose the following bases \bea
\psi^{-}&=&\left( \begin{array}{ccccccccc} l_{1} & l_{2} & l_{3}
&\tilde{\mathcal{W}}^{-} &\tilde{\mathcal{Y}}^{-}&
\tilde{\rho}^{\prime -}_{1}& \tilde{\rho}^{\prime -}_{2}&
\tilde{\chi}^{-}
\end{array}
\right)^{T},  \crn \psi^{+}&=&\left( \begin{array}{ccccccccc}
l_{1}^{c} & l_{2}^{c} & l_{3}^{c} &&\tilde{\mathcal{W}}^{+}
&\tilde{\mathcal{Y}}^{+}& \tilde{\rho}^{+}_{1}&
\tilde{\rho}^{+}_{2}&\tilde{\chi}^{\prime +}
\end{array}
\right)^{T}, \eea and define \be
\Psi^{\pm}=\left(\begin{array}{cc} \psi^{+} & \psi^{-}
\end{array}\right)^{T}.
\ee With these definitions, the mass term is written in the form
\be - \frac{1}{2}( \Psi^{\pm})^{T}Y^{\pm}\Psi^{\pm}+ H.c., \ee
where \be Y^{\pm}= \left( \begin{array}{cc}
0 & X^{T} \\
X & 0
\end{array}\right).
\ee
Here the $X$ matrix is given by
\bea X= \left(%
\begin{array}{cccccccc}
  \fr{\la_{111}}{3 \sqrt{2}}v' & \fr{\la_{121}}{3 \sqrt{2}}v' &
  \fr{\la_{131}}{3 \sqrt{2}}v' & 0 & 0 & 0 & 0 & 0  \\
  \fr{\la_{112}}{3 \sqrt{2}}v' & \fr{\la_{122}}{3 \sqrt{2}}v' &
  \fr{\la_{132}}{3 \sqrt{2}}v' & 0 & 0 & 0 & 0 & 0  \\
  \fr{\la_{113}}{3 \sqrt{2}}v' & \fr{\la_{123}}{3 \sqrt{2}}v' &
  \fr{\la_{133}}{3 \sqrt{2}}v' & 0 & 0 & 0 & 0 & 0  \\
  0 & 0 & 0 &{\mathcal{M}}_{W} & 0 & \fr{gv'}{\sqrt{2}} & 0 & \fr{gu}{\sqrt{2}}  \\
  0 &0 & 0 & 0 &{\mathcal{M}}_{Y} & 0 & \fr{gv'}{\sqrt{2}} & \fr{gw}{\sqrt{2}}  \\
  \fr{\la_{21}}{3 \sqrt{2}}w & \fr{\la_{22}}{3 \sqrt{2}}w &
  \fr{\la_{23}}{3 \sqrt{2}}w & \fr{gv}{\sqrt{2}} & 0 & \mu_{\rho} &
  0 & 0 \\
  \fr{\la_{21}}{3 \sqrt{2}}u & \fr{\la_{22}}{3 \sqrt{2}}u &
  \fr{\la_{23}}{3 \sqrt{2}}u & 0 & \fr{gv}{\sqrt{2}} & 0 & \mu_{\rho} &
  0  \\
  \fr{1}{2}\mu_{01} & \fr{1}{2}\mu_{02} & \fr{1}{2}\mu_{03} &
  \fr{gw}{\sqrt{2}} & \fr{gw'}{\sqrt{2}} & 0 &
   0 & \mu_{\chi} \\
\end{array}
\right).\nn
\label{genraleletronmassmatrix}
\eea
 The chargino mass matrix $Y^\pm$ is diagonalized by
using two unitary matrices, $D$ and $E$, defined by \bea \tilde{
\chi}^{+}_{i}=D_{ij} \Psi^{+}_{j}, \,\ \tilde{
\chi}^{-}_{i}=E_{ij} \Psi^{-}_{j}, \,\ i,j=1, \cdots , 8.
\label{2sc} \eea
 The characteristic equation for the  matrix
$Y^{\pm}$  is \bea \det (Y^{\pm}- \lambda I)= \det \left[ \left(
\begin{array}{cc}
- \lambda & X^{T} \\
X  &- \lambda
\end{array} \right) \right]= \det( \lambda^{2}-X^{T} \cdot X).
\label{propmat1} \eea
  Since $X^{T} \cdot X$ is a symmetric
matrix, $\lambda^2$ must be real and positive because $Y^{\pm}$ is
also symmetric. In order to obtain eigenvalues,  one only have to
calculate $X^{T} \cdot X$.  The diagonal mass
matrix can be written as \be M_{SCM}=E^{*}XD^{-1}. \label{m1} \ee
 To determine $E$ and $D$, it is useful the following observation
\be M^{2}_{SCM}=DX^T \cdot XD^{-1}=E^{*}X \cdot X^T(E^{*})^{-1}.
\label{m2} \ee It means that $D$ diagonalizes $X^{T} \cdot X$,
while $E^{*}$ diagonalizes $X \cdot X^{T}$. In this case we can
define the following Dirac spinors: \bea \Psi(\tilde{
\chi}^{+}_{i})= \left(
\begin{array}{cc}
             \tilde{ \chi}^{+}_{i} &
         \bar{ \tilde{\chi}}^{-}_{i}
\end{array} \right)^T, \,\
\Psi^{c}(\tilde{ \chi}^{-}_{i})= \left( \begin{array}{cc}
             \tilde{ \chi}^{-}_{i} &
         \bar{ \tilde{\chi}}^{+}_{i}
\end{array} \right)^T.
\label{emasssim} \eea where $\tilde{ \chi}^{+}_{i}$ is the
particle and $\tilde{ \chi}^{-}_{i}$ is the
anti-particle~\cite{mcr}.

 Using the values given in Eqs.(\ref{para},\ref{vevs}), the
eigenvalues of the charged fermion matrix given at
Eq.(\ref{genraleletronmassmatrix}) are obtained as
\begin{eqnarray}
m_{e}&=&5 \times 10^{-4}\ \mathrm{GeV}, \ m_{\mu}=0,104\
\mathrm{GeV}, \ m_{\tau}=1,179\ \mathrm{GeV}, \crn
m_{\tilde{\chi}^{\pm}_{1}} &= &69,09\ \mathrm{GeV}, \
m_{\tilde{\chi}^{\pm}_{2}}=522,92\ \mathrm{GeV}, \nonumber \\
m_{\tilde{\chi}^{\pm}_{3}}&=&600,22\ \mathrm{GeV}, \
m_{\tilde{\chi}^{\pm}_{4}}=732,64\ \mathrm{GeV}, \
m_{\tilde{\chi}^{\pm}_{5}}=1168\ \mathrm{GeV}.
\label{completechargedleptonsmassespectrum}
\end{eqnarray}

It is easily to see that the mass matrix is divided in two
sectors: one heavy which is identified as the charginos and has
been studied in our previous work \cite{Dong:2007qc}. Another one
light which is identified as the usual leptons.

On the other hand, if we take the mass matrix from our work and
using the same values to the parameters  we get masses of the
charged leptons:
\begin{eqnarray}
m_{e}=5 \times 10^{-4}\ \mathrm{GeV}, \
m_{\mu}=0,119\ \mathrm{GeV}, \
m_{\tau}=1,249\ \mathrm{GeV}.
\end{eqnarray}

Now using the mass matrix to the charginos  given in
\cite{Huong:2008ww}
we get
\begin{eqnarray}
&& m_{\tilde{\chi}_{1}}=69,05\ \mathrm{GeV}, \
m_{\tilde{\chi}_{2}}=522,78\ \mathrm{GeV}, \
m_{\tilde{\chi}_{3}}=600,01\ \mathrm{GeV}, \crn &&
m_{\tilde{\chi}_{4}}=600,42\ \mathrm{GeV}, \
m_{\tilde{\chi}_{5}}=1168\ \mathrm{GeV}.
\mathrm{GeV}.
\end{eqnarray}

In this  case, the new elements put down   masses of the muon and
tauon  and  at the same time remove the mass degeneracy between
$\chi_{3}$ and $\chi_{4}$. Anyway we can see that the mass matrix
in the charged fermion  sector is basically divided in two
sectors: one giving masses of the usual known leptons and another
one giving  masses of the new charginos.

\section{Nucleon decay in the SUSYECO331}
\label{pmssm}

Let us remind  that in the
MSSM~\cite{Fayet:1974pd,Fayet:1976et,Fayet:1977yc,fayet78rp,mssm},
the $R$-parity violating terms in superpotential  are given by
$W_{2RV}+W_{3RV}+ \overline{W}_{2RV}+\overline{W}_{3RV}$, where
\begin{eqnarray}
W_{2RV}&=&\mu_{0a}\epsilon \hat{L}_a\hat{H}_2,\nonumber \\
W_{3RV}&=&\lambda_{abc}\epsilon\hat{L}_{a}\hat{L}_{b}\hat{l}^{c}_{c}+
\lambda^{\prime}_{iaj}\epsilon\hat{Q}_{i}\hat{L}_{a}\hat{d}^{c}_{j}+
\lambda^{\prime\prime}_{ijk}\hat{d}^{c}_{i}\hat{u}^{c}_{j}\hat{d}^{c}_{k}.
\label{mssmrpv}
\end{eqnarray}
Here we have suppressed $SU(2)$ indices, $\epsilon$ is the
antisymmetric $SU(2)$ tensor. Above, and below in the following,
the subindices $a,b,c$ run over the lepton generations
$e,\mu,\tau$ but a superscript $^c$ indicates charge conjugation;
$i, j, k=1, 2, 3$ denote quark generations.

Note that the $\lambda^{\prime}$-coupling in the MSSM is similar
to the $\xi_{7}$-coupling in the most general form of the
superpotential in the SUSYECO331, which is  given in
Eq.(\ref{w3}), and the coupling $\lambda^{\prime\prime}$ is
similar to the coupling $\xi_{3}$.

\begin{figure}[ht]
\begin{center}
\vglue -0.009cm 
\mbox{\epsfig{file=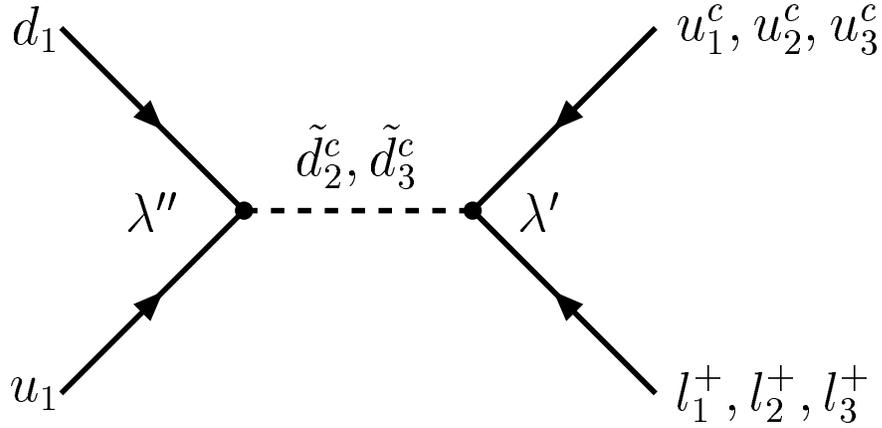,width=0.7\textwidth,angle=0}}
\end{center}
\caption{Proton decay into charged leptons in the MSSM
and in SUSYECO331 (with $\lambda^{\prime \prime}
\rightarrow \xi_{3}$ and $\lambda^{\prime} \rightarrow \xi_{7}$).}
\label{f1}
\end{figure}

\begin{figure}[ht]
\begin{center}
\vglue -0.009cm 
\mbox{\epsfig{file=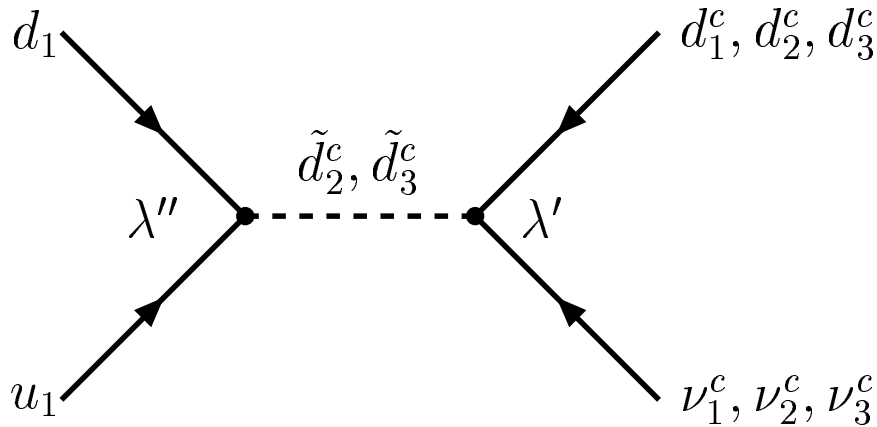,width=0.7\textwidth,angle=0}}
\end{center}
\caption{Proton decay into antineutrinos in the  MSSM and in
SUSYECO331 (with $\lambda^{\prime \prime} \rightarrow \xi_{3}$ and
$\lambda^{\prime} \rightarrow \xi_{7}$).} \label{f2}
\end{figure}

From  Eq. (\ref{mssmrpv}), we can obtain the $B$-violating Yukawa
couplings as follows
\begin{eqnarray}
\lambda^{\prime \prime}_{ijk}
\bar{d}_{iR}u^{c}_{jL}\tilde{d}^{c}_{k}+H.c.
\label{lpp}
\end{eqnarray}
The interactions among lepton, quark and squark are given by
\begin{eqnarray}
\lambda^{\prime}_{iaj}\left( \bar{d}^{c}_{iR}\nu_{aL}-
\bar{u}^{c}_{iR}l_{aL} \right) \tilde{d}_{j}+H.c. \,\ .
\label{lpmssm}
\end{eqnarray}
Taking into account  Eqs. (\ref{lpp},\ref{lpmssm}), we can draw
two Feynmann diagrams describing  the proton decay, which are
shown in the Fig.\ref{f1} and Fig.\ref{f2}. The Fig.\ref{f1}
describes the proton decay into charged leptons.
At the lowest
 approximation, there is  no mixing in
the quark, neutrino and squark sectors. It means that $u_{1}
\equiv u$, $u_{2} \equiv c$ and $u_{3} \equiv t$ and so on. The
proton could decay into $p \to \pi^{0}e^{+}$,  $p \to
\bar{D}^{0} \mu^{+}$ and $p \to (u_1 t^{c}) \tau^{+}$, but the
last two modes are forbidden kinematically.

The analysis presented above shows that the proton can decay only
in $p \to \pi^{0}e^{+}$. On  dimensional grounds, we estimate
\begin{eqnarray}
\Gamma (p \to \pi^{0} e^{+})\approx
\frac{\alpha(\lambda^{\prime}_{11k}) \alpha(\lambda^{\prime
\prime}_{11k})}{ m^4_{{\tilde d_{k}}}}{M_{proton}^5},
\end{eqnarray}
where $\alpha(\lambda)=\lambda^2/(4\pi)$. Giving  $\tau(p \to e
\pi)
>1.6 \times 10^{33}$ years \cite{pdg} and taking
$m_{\tilde{d}_k} \sim  {\cal O}(1 \textrm{TeV})$, we obtain
\begin{equation}
\lambda^{\prime}_{11k}\lambda^{\prime \prime}_{11k}\; < \; 5.29
\times 10^{-26}. \label{proton}
\end{equation}
It is consistent with the limits presented in \cite{drei}. For a
more detailed calculation see \cite{sher,hin}.
Other decay modes, where the proton decay into antineutrino, have
been  considered in \cite{smi}. The mentioned decay modes are $p
\to \pi^{+}\bar{\nu_{e}}$, $p \to K^{+}\bar{\nu_{\mu}}$, $p \to
B^{+}\bar{\nu_{\tau}}$\footnote{Again, we are bare mixing in the
quarks, neutrinos and squarks sectors}. In these cases, we get the
same numerical results as presented in Eq.(\ref{proton}).

The bound presented in Eq.(\ref{proton}) is so strict. So
the  natural explanation  only is that at least one of the
couplings has to be zero. In order to avoid that the simplest way
is to impose the $R$-symmetry.
This leads to avoid the proton decay.

In our superpotential given at Eq.(\ref{neutrinomassesseco331}),
we allow only interactions that violate $L$-number. Therefore the
proton is stable at tree-level. However it is
not only forbidden the dangerous processes of proton decay but
also forbidden the neutron-antineutron oscillation. This
oscillation was studied in detail  in
\cite{dress,tata,barbier,moreau}.

 In \cite{susy3312,Dong:2006vk}, in the framework of the supersymmetric 3-3-1
model with right handed neutrinos, the R-parity violating
interaction was applied for instability of the nucleon. The result
is consistent with that of the present article (noting that in Ref
\cite{susy3312}, the authors have taken a lower limit of the
proton lifetime equal to  $10^{32}\  years$ \textbf{)}.

\section{Neutrinoless double beta decay  in   SUSYECO331}

\begin{figure}[ht]
\begin{center}
\vglue -0.009cm 
\mbox{\epsfig{file=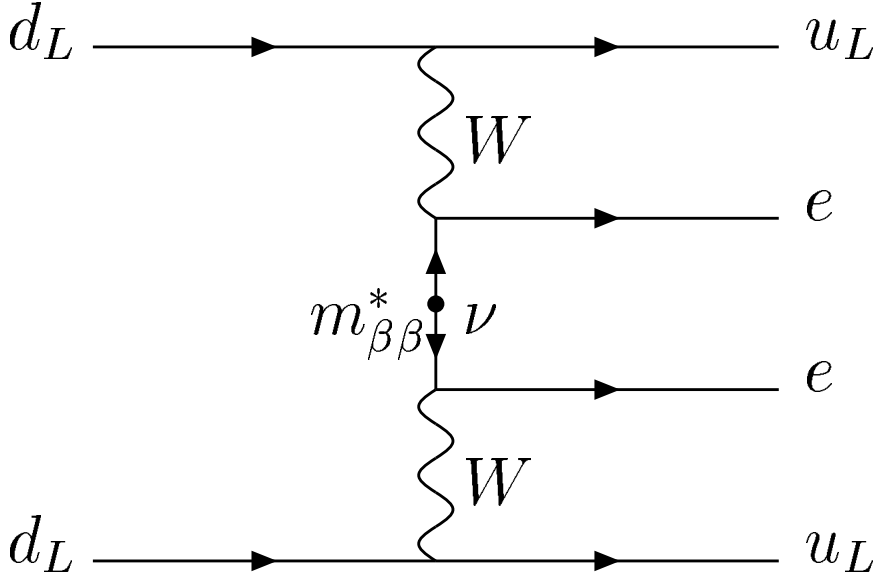,width=0.7\textwidth,angle=0}}
\end{center}
\caption{Neutrinoless double beta  decay in  SM with massive
neutrinos.} \label{ddbsnmp}
\end{figure}

Neutrinoless double beta ($0\nu\beta\beta$) decay is a sensitive
probe of physics beyond the standard model, see
Fig.(\ref{ddbsnmp}), since it violates the conservation of lepton
number \cite{hdmo94,baudis97,heidel01,expbb}. Such experimental
results if available will cast stringent constraints on new
physics.
The nucleon level process of $0\nu\beta\beta$ decay, $ n +n \to
p+p + e^- + e^-$, can be obtained via the lepton number violating
subprocess, $ d+d \to u +u +e+e$. For experiments, the parent
nucleus $ A_Z$ can decay to a daughter nucleus $A_{Z+2} $ via the
two steps of virtual transitions, $A_Z \to A _{Z+1} +\beta ^- \to
A_{Z+2} +\beta ^- +\beta ^- $. It is to be noted that only the
ordinary double beta reactions have been experimentally observed,
while the active searches for the
 $0\nu\beta\beta$ reactions are actually get with null results.
The experimental measurements of $0\nu\beta\beta$ decay are
currently accounted for even-even heavy nuclei such as $^{48}Ca
\to \ ^{48} Ti ,\ ^{76} Ge \to
 \   ^{76} Se ,\ ^{82} Se \to  \   ^{82} Kr , \ ^{100} Mo
 \to  \   ^{100} Ru ,\ ^{128} Te \to  \
^{128}Xe $. The most stringent bounds have been given by the
Moscow-Heidelberg collaboration~\cite{baudis97,heidel01} on the
$0\nu\beta\beta$ decay half-life of $^{76} Ge $ as $T_{ud}
> [1.1\ \times 10 ^ {25},\ 1.5\ \times 10 ^ {25}] $ yr,
respectively. We can find in Ref. \cite{arxivnote} for a general
review, and Ref.~\cite{expbb} for a summary of the experimental
results as well as the future projects.

\begin{figure}[ht]
\begin{center}
\vglue -0.009cm 
\mbox{\epsfig{file=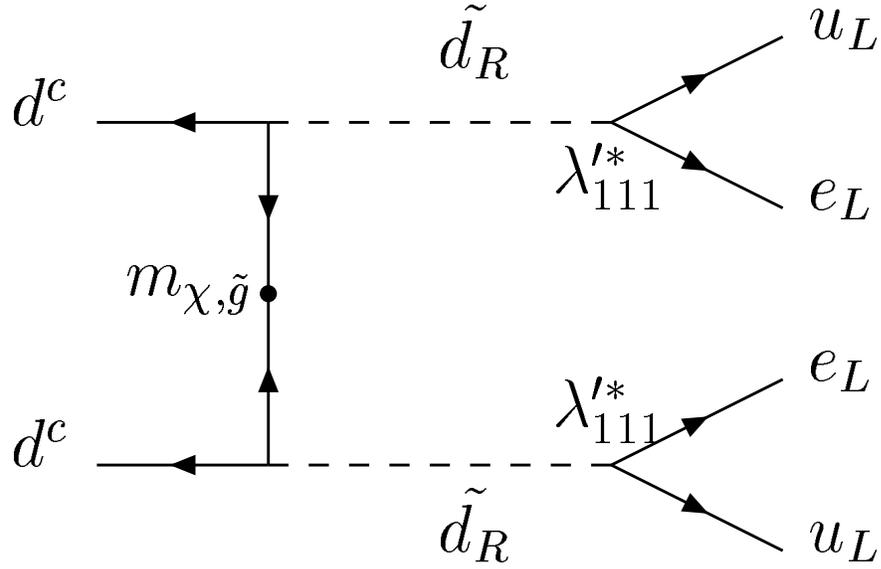,width=0.7\textwidth,angle=0}}
\end{center}
\caption{Neutrinoless double beta ($0\nu\beta\beta$) decay in the
MSSM with massive neutrinos and in SUSYECO331 (with
$\lambda^{\prime} \rightarrow \xi_{7}$).} \label{ddbsnmpsm}
\end{figure}

Particularly, for the conventional $0\nu\beta\beta$-decay with
massive Majorana neutrino exchange it implies an upper bound on
the neutrino mass below $1$ eV. The supersymmetric mechanism of
$0\nu\beta\beta$ decay was first suggested by Mohapatra
\cite{Mohapatra} and further studied in Refs.
\cite{Vergados,HKK1}. In Ref. \cite{HKK2}, the $R$ parity
violating Yukawa coupling of the first generation is strongly
bounded by $\lambda'_{111} \leq 3.9\cdot 10^{-4}$ due to the
gluino exchange $0\nu\beta\beta$-decay. Babu and Mohapatra
\cite{BM} have latter implemented another contribution comparable
with that via the gluino exchange. This set stringent bounds on
the products of $R$ parity violating Yukawa couplings
$\lambda'_{11i}\lambda'_{1i1}$ of $i$th generation index, see
Fig.(\ref{ddbsnmpsm}).
The constraints on $\lambda'_{111}$ coming from the
$0\nu\beta\beta$ half--life limit have been calculated
\cite{hir96,hir96c}
\begin{equation}
\lambda^{'}_{111} \leq 3.9 \cdot 10^{-4} \Big(\frac
{m_{\tilde{q}}}{100\  \textrm{GeV}} \Big)^2 \Big(\frac
{m_{\tilde{g}}}{100\  \textrm{GeV}} \Big)^{\frac{1}{2}}
\end{equation}
with the assumption $m_{\tilde{d_R}} \simeq m_{\tilde{u}_L}$ and
$m_{\tilde{q}}$, $m_{\tilde{g}}$ being squark and gluino masses,
respectively.

We find further \cite{hir96} \bea
\lambda_{113}^{'}\lambda_{131}^{'} &\leq & 1.1 \cdot 10^{-7},\\
\lambda_{112}^{'}\lambda_{121}^{'}&\leq & 3.2 \cdot 10^{-6}. \eea
The analyses presented in the MSSM  to the $0\nu\beta\beta$-decay
are still hold in the SUSYECO331 model, because the
$\lambda^{\prime}$ is allowed in our superpotential as shown at
Eq.(\ref{neutrinomassesseco331}).

\section{Conclusion}
\label{conclusion}

In this paper we have presented new  $R$-symmetry  for the
supersymmetric economical $\mbox{SU}(3)_{C}\otimes
\mbox{SU}(3)_{L} \otimes \mbox{U}(1)_{X}$  mode and studied
neutrino mass by implication for the obtained $R$-parity. The
neutrino mass spectrum is affected by the chosen R-parity. By
imposing the $R$-parity, namely:  $
\hbox{R-parity}=(-1)^{2S}(-1)^{3({\cal B}+{\cal L})}$, the
neutrino mass spectrum at the tree level contains two massless.

We remind that in the non-supersymmetric economical 3-3-1 model,
to give neutrinos a correct mass pattern, we have to introduce new
mass scale around the GUT scale. The same situation happens in the
supersymmetric version and the above puzzle can be solved with the
help of the inflaton having mass in the range of the GUT scale. In
this paper, we have showed that by the chosen $R$ parity and the
set of parameters, the pseudo-Dirac neutrino mass is available.

We have found  that the other $R$-parity given in
(\ref{Rnew}\textbf{)} leads to the interference between the
neutrino and neutralino mass matrices.
 Because of this interference
mass matrix, all neutrino gain mass only just at the tree level
and the interference mass matrix does not  much affect on the
neutralino mass spectrum. In the charged lepton sector, with the
new $R$ parity,
there is also an interference mass matrix between the usual
leptons and charginos. By taking the numerical, we show that the
charged sector is basically divided in two distinct sectors: one
giving the usual known leptons and the other ones given the new
charginos. If we  ignore the interference charged lepton mass
matrix, the chargino mass spectrum is degenerated. This
degenerated mass spectrum is removed by imposing the new R-parity.

\textbf{T}he new $R$-parity not only provides a simple mechanism for the
mass generation of the neutrinos but also gives  some lepton
flavor violating  interactions at the tree level. This will play
some important phenomenology in our model such as the proton's
stability, forbiddance of the neutron-antineutron oscillation and
neutrinoless double beta decay.

\section*{Acknowledgments}
M. C. R. is grateful to Conselho Nacional de Desenvolvimento
Cient\'\i fico e Tecnol\'ogico (CNPq) under the processes
309564/2006-9 for supporting his work, he also would like to thank
  Vietnam Academy of Science and Technology
for the nice hospitality, warm atmosphere during his stay at
Institute of Physics to do this work.
 This work was supported in part by the National Foundation for Science
and Technology Development (NAFOSTED)  under grant  No:
103.01.16.09.

\appendix

\section{Elements of  neutrino mass matrix  in SUSYECO331}
\label{neumatel}

The elements of $Y^0$ presented in Eq. (\ref{mc1}) is given by
Eq.(\ref{neumatr}) where
\begin{equation}
M_{\nu} = \left(
\begin{array}{cccccc}
0 & 0 & 0 & 0 & G_{12} & G_{13} \\
0 & 0 & 0 & G_{21} & 0 & G_{23} \\
0 & 0 & 0 & G_{31} & G_{32} & 0 \\
0 & G_{21} & G_{31} & 0 & 0 & 0  \\
G_{12} & 0 & G_{32} & 0 & 0 & 0  \\
G_{13} & G_{23} & 0 & 0 & 0 & 0  \\
\end{array}
\right) \label{neumatr1}
\end{equation}
and
\begin{equation}
G_{ab}=\frac{v}{3 \sqrt{2}} \left( \lambda_{3ab}- \lambda_{3ba}
\right),
\end{equation}
while
\begin{equation}
M_{\nu N}= \left(
\begin{array}{ccccccccccc}
0 & \frac{\mu_{01}}{2} & 0 & \frac{\lambda_{21}}{3 \sqrt{2}}v &
\frac{\lambda_{21}}{3 \sqrt{2}}w & 0 & 0 & 0 & 0 & 0 & 0 \\
0 & \frac{\mu_{02}}{2} & 0 & \frac{\lambda_{22}}{3 \sqrt{2}}v &
\frac{\lambda_{22}}{3 \sqrt{2}}w & 0 & 0 & 0 & 0 & 0 & 0 \\
0 & \frac{\mu_{03}}{2} & 0 & \frac{\lambda_{23}}{3 \sqrt{2}}v &
\frac{\lambda_{23}}{3 \sqrt{2}}w & 0 & 0 & 0 & 0 & 0 & 0 \\
0 &- \frac{\lambda_{21}}{3 \sqrt{2}}v & 0 & \frac{\mu_{01}}{2} &-
\frac{\lambda_{21}}{3 \sqrt{2}}w & 0 & 0 & 0 & 0 & 0 & 0 \\
0 &- \frac{\lambda_{22}}{3 \sqrt{2}}v & 0 & \frac{\mu_{02}}{2} &-
\frac{\lambda_{22}}{3 \sqrt{2}}w & 0 & 0 & 0 & 0 & 0 & 0 \\
0 &- \frac{\lambda_{23}}{3 \sqrt{2}}v & 0 & \frac{\mu_{03}}{2} &-
\frac{\lambda_{23}}{3 \sqrt{2}}w & 0 & 0 & 0 & 0 & 0 & 0
\end{array}
\right)\label{neumatr2}
\end{equation}
and $M_{N}$ is presented in \cite{Huong:2008ww}.

\end{document}